# Protein structure, activity and thermal stability within nanoscopic compartments

Denis Cecchin[1,2], Luca Chierico[1], Xiaohe Tian[1,4,5], Katharina Kluthe[1,6], Alessandro Poma[1,2], Lorena Ruiz-Pérez[1,2,3], Caterina LoPresti[7], and Giuseppe Battaglia[1,2,3,8,9*]

*[1]Department of Chemistry, [2]Institute for the Physics of Living Systems, [3]EPSRC/Jeol Centre for Liquid Phase Electron Microscopy, University College London, London, UK. [4]Department of Chemistry [5]School of Life Science, Anhui University, Hefei, P. R. China. [6]Max Planck Institute for Polymer Research, Mainz, Germany, [7]Pharmaceutical Development Early Phase Biologics, Novartis, Basel, Switzerland, [8]Institute for Bioengineering of Catalonia (IBEC), The Barcelona Institute for Science and Technology (BIST), Barcelona, Spain. [9]Institució Catalana de Recerca i Estudis Avançats (ICREA), Barcelona, Spain*

## Abstract

We report that protein confinement within nanoscopic vesicular compartments corresponds to a liquid-liquid phase transition with the protein/water within vesicle lumen interacting very differently than in bulk. We show this effect leads to considerable structural changes on the proteins with evidence suggesting non-alpha helical conformations. Most importantly both aspects lead to a significant improvement on protein stability against thermal denaturation up to 95ºC at neutral pH, with little or no evidence of unfolding or reduced enzymatic activity. The latter parameter does indeed exhibit an increase after thermal cycling. Our results suggest that nanoscopic confinement is a promising new avenue for the enhanced long-term storage of proteins. Moreover, our investigations have potentially important implications for the origin of life, since such compartmentalization may well have been critical for ensuring the preservation of primordial functional proteins under relatively harsh conditions, thus playing a key role in the subsequent emergence of primitive life forms.

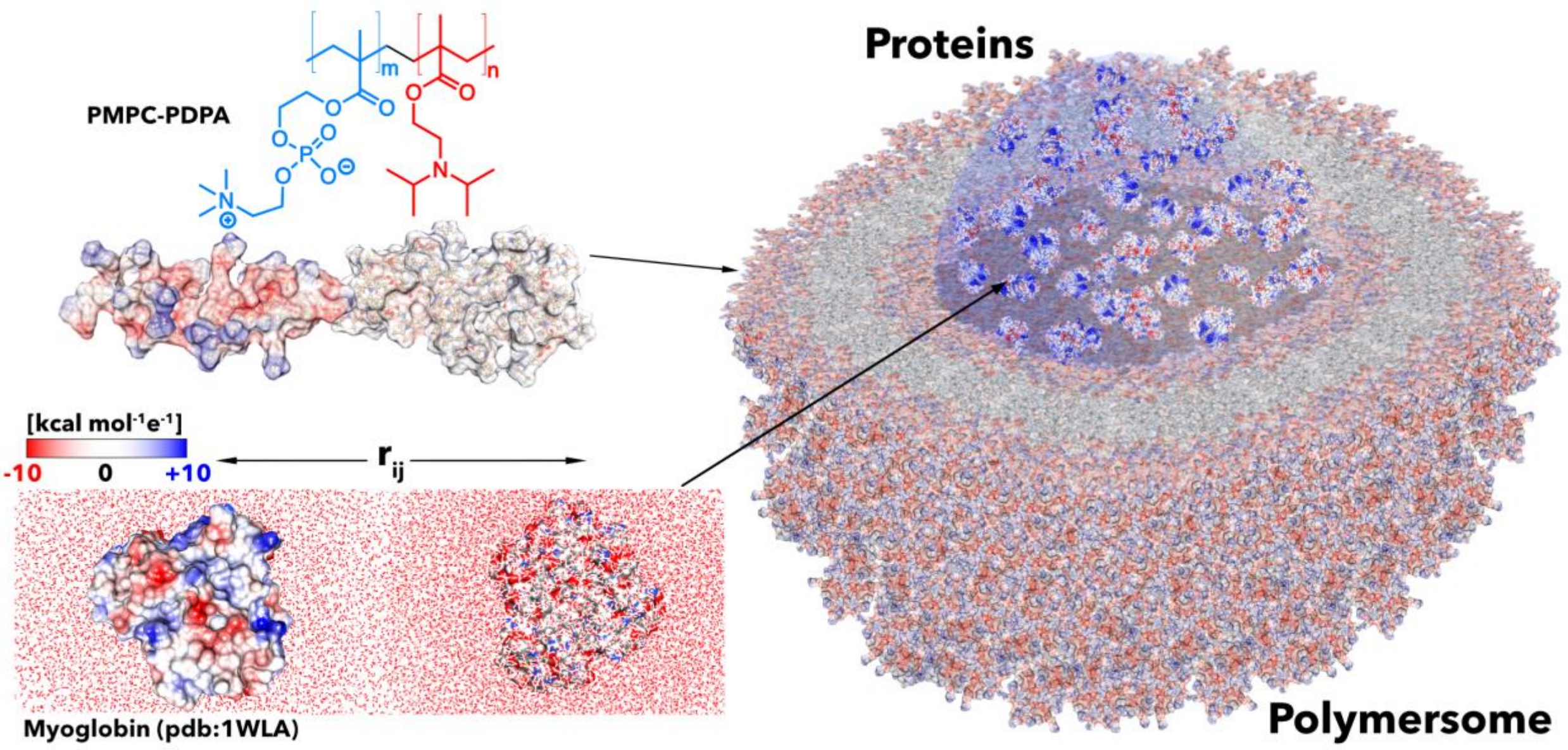

**Fig.1** Schematic representation of PMPC-PDPA polymersomes encapsulating proteins on the polymersomes' core

## Introduction

Proteins result from the sequence-controlled polymerization of amino acids into long chains that fold into functional three-dimensional structures (1). The understanding of the folding process, as well as the final 3D structure, has been one of the most productive areas where chemistry, physics and biology have merged and contributed to the understanding of biological complexity as well as feeding most drug design efforts (2). Today we have very little understanding of how protein folding, structure and function are regulated into complex biological milieus. In living cells, compartments confine proteins to local concentrations that can reach up to 40% in volume creating highly crowded, confined and saturated aqueous solutions (3, 4). We know today that crowding affects the protein, structure, enzymatic conversion, and diffusion, as well as the diffusion of any other small molecules, dispersed in the remaining space (5, 6). Most importantly, crowding restricts that the water that baths all the components with consequently altered properties compared to free bulk water (7, 8). The study of protein structure and function within confined spaces has significant ramifications across several areas ranging from protein therapeutics to the food industry (9)(10).

The structure of proteins is the result of a delicate balance between various supramolecular forces, including the hydrophobic effect, hydrogen bonding, sulphur bridge formation, electrostatic and aromatic interactions (1, 11). These supramolecular forces are normally represented by energy landscapes whereby the protein folded conformation corresponds to the absolute minimum energy (1). Today we have solved the structure of over 150,000 proteins (12), and most of these studies have been done isolating the protein and studying it under dilute conditions often neglecting effects such as hydration, protein-protein interactions, etc. (8). Over the last two decades, macromolecular crowding has been studied by the addition of high concentrations of various macromolecules to aqueous protein solutions, such as poly(ethylene oxide), dextran, hemoglobin or defatted albumin (13). While these studies have demonstrated that crowding favors protein folding (14), little or no effect on the thermal stability of proteins has been observed (15).

In contrast, proteins confined within silica gels (16-18), polymeric gels (19) or mesoporous silicates (20) exhibit enhanced thermal stability due to confinement effects (21). The latter is quite extreme in all such experiments with the available volume being very close to that of a single folded protein, suggesting minimal hydration. Under such strong geometric constraints, there is almost no space available for proteins to unfold even if chemical instability were to ensure (21). However, such a stabilizing effect is only possible when water is free to diffuse in and out of the confined volume (21). While these studies provide interesting insights regarding protein dynamics, they are often limited by the strong interaction between the confinement/crowding agent and the protein, leading to 'unnatural' denaturation driven by the agent itself (17). Such artificial conditions do not adequately represent those generally found within the cell interior, where hydration and protein/protein interactions play a significant role in controlling the (un)folding dynamics. Although the whole cell is micrometer in size, its interior is often compartmentalized with membrane-bound volumes ranging from tens to hundreds of nanometers (22). The presence of the membrane adds an interface to the water pool and consequently further affects the water properties. The membrane hydration itself is not restricted to the membrane/water interface but extends of few nanometers into the cytosol making it considerably different from bulk water (23).

In the present work, we study the simultaneous effect of confinement on protein stability and structure using block copolymer vesicles also known as polymersomes (24). Polymersomes comprise membrane-enclosed nanoscopic compartments produced by the self-assembly of amphiphilic diblock copolymers in aqueous solution (25). Polymersomes morphology and supramolecular nature are very similar to those found on

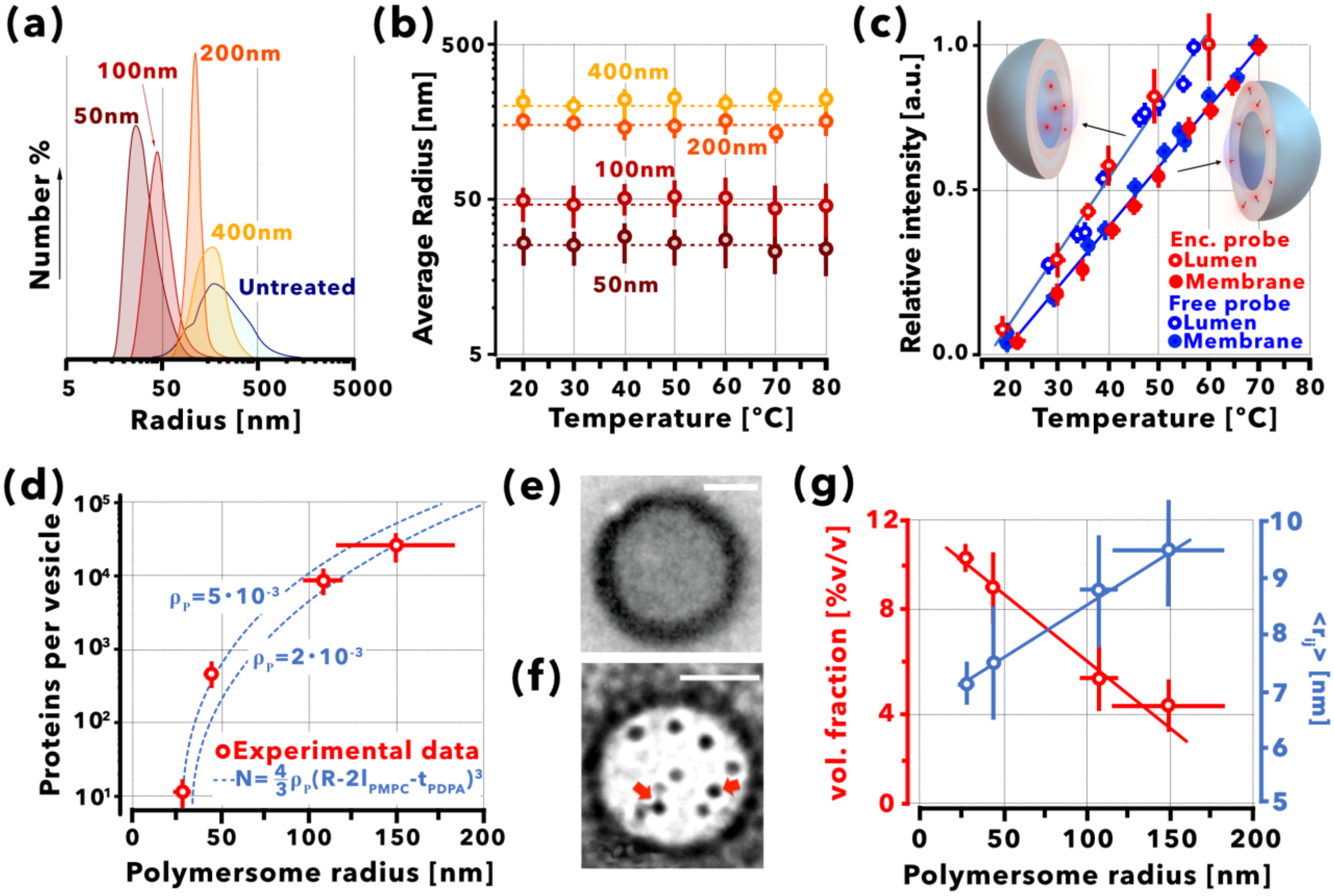

**Fig.2 (a)** Size distributions measured by dynamic light scattering (DLS) of PMPC-PDPA polymersomes formed by film rehydration and extruded through polycarbonate membranes with pore size of 50, 100, 200 and 400nm. **(b)** Average hydrodynamic radius measure by DLS of the different polymersomes populations obtained by extrusion as a function of the temperature (n = 3; error bars = ± SD). **(c)** Relative fluorescence intensity of free and encapsulated Rhodamine B - $PMPC_{25}$ and Rhodoamine B octadecyl ester as a function of the temperature. Note free Rhodoamine B octadecyl ester fluorescence was measured in octanol (n = 3; error bars = ± SD). **(d)** Number of myoglobin protein per polymersomes as a function of the polymersomes radius (n = 3; error bars = ± SD). Transmission electron micrographs of empty an empty **(e)** and loaded with 5nm gold nanoparticles conjugated IgG **(f)** PMPC-PDPA polymersomes. Note the scale bar is 50nm. **(g)** The lumen myoglobin protein volume fraction and the corresponding mean inter-particle distance $<r_{ij}>$ as a function of the polymersome radius (n = 3; error bars = ± SD).

natural cell organelles. Polymersomes are however much more robust structures lipid ones and allow for an accurate control over both structural and functional parameters (25). Polymersome systems have been recently proposed as effective carriers for the delivery of drugs, nucleic acids and proteins (24, 26). In particular, we have demonstrated that pH-sensitive polymersomes based on poly(2-(methacryloyloxy)ethyl phosphorylcholine)-poly(2-(diisopropylamino)ethyl methacrylate) (PMPC-PDPA) can deliver payloads within live cells with no detrimental effect to cell viability (27). Here we demonstrate the effective encapsulation of myoglobin within PMPC-PDPA polymersomes and show how such nanoscopic confinement allows for protein protection.

## Results and discussion

**Polymersome preparation and characterization.** PMPC-PDPA polymersomes (**Fig.1**) comprise four critical properties for studying protein encapsulation: (*i*) the PMPC block is a highly hydrated water-soluble polymer that is strongly protein-repellent (28, 29); (*ii*) The main membrane/water interface hydration is controlled by the phosphorylcholine group the most abundant hydrophilic head expressed by natural phospholipids;

(*iii*) PMPC-PDPA polymersomes are stable at high temperature (30, 31) opposite to phospholipids which lose stability at around 60ºC (32, 33); and (*iv*) the pH-sensitive nature of the PDPA block, allows for the efficient and reversible encapsulation/release of large macromolecules (27, 34, 35). PMPC polymersomes were prepared using film rehydration techniques, purified by centrifugation and size exclusion chromatography (36) and extruded through porous polycarbonate membranes with pore diameters of 50, 100, 200 or 400 nm (37, 38). Polymersome size distributions determined using dynamic light scattering (DLS) is plotted in **Fig. 2a** for dispersions extruded through the various polycarbonate membranes alongside the untreated sample. We further confirmed the colloidal stability as a function of the temperature **Fig. 2b** showing that no notable difference in the hydrodynamic radius was detected for all the different size dispersions. In addition to this, we encapsulated two different Rhodamine B probes, one amphiphilic probe housed within the membrane: Rhodamine B octadecyl ester, and one hydrophilic probe housed within vesicle lumen: Rhodamine B-$PMPC_{25}$ polymer. Rhodamine B fluorescence intensity is strongly affected by the temperature (39) and allows precise measurement of thermal changes as we show in the graph in **Fig. 2c**, we observed no difference in the temperature monitored when the probe was located either in the polymersome membrane or the lumen. We thus conclude that PMPC-PDPA polymersome membrane does not act as an insulator and the temperature within its lumen and membrane is equal to the temperature in the bulk solution.

**Protein encapsulation.** We have proved that PMPC-PDPA polymersomes can encapsulate several types of proteins including immunoglobulin G (IgG) (40, 41), albumin (41, 42), myoglobin (41, 43), trypsin (44), catalase (45), and glucose oxidase (45). We have also shown that proteins do not interact with the PMPC stabilized surface of the polymersomes and hence protein encapsulation is due to actual entrapment within the lumen (46). Protein encapsulation can be achieved either during polymersome preparation (35, 40) or post polymersome formation by controlled electroporation (41). We encapsulated both IgG and myoglobin by electroporation where sequential pulses of alternative electric currents are used to induce temporary poration in the polymersomes membrane. The temporary membrane disruption allows for the proteins dissolved in bulk to enter the lumen (44). This latter method allows to prepare and purify polymersomes to obtain controlled size dispersions. As shown in **Fig. 2d**, we encapsulated myoglobin within different size polymersomes and measured the number of proteins encapsulated, $N_P$, as a function of the polymersome radius, $R_P$. Using simple geometrical considerations, we can write the equation

$$N_P = \frac{4}{3}\pi\rho_P(R_P - 2l_{PMPC} - t_{PDPA})^3 \quad (1),$$

where $\rho_P$ is the protein particle density within the polymersome lumen expressed as the number of molecules per volume, $l_{PMPC}$ is the $PMPC_{25}$ chain length estimated from simulations (31, 47) to be 6nm, and $t_{PDPA}$ is the PDPA membrane thickness which we have measured to be 7nm (30, 31). We use equation 1 to fit the data, and as shown in **Fig. 2d**, the experimental data fall within two particle densities $\rho_P$ of $5 \cdot 10^{-3}$ and $2 \cdot 10^{-3}$ $nm^{-3}$ respectively suggesting that the smaller vesicles are more efficient in encapsulating proteins. Lumen encapsulation is further confirmed by transmission electron microscopy (TEM) using 5nm gold nanoparticles conjugated to IgG (GNP-IgG) as a model protein. As shown in **Fig. 2e**, the empty polymersomes show the typical vesicular geometry while when loaded with the GNP-IgG the gold is visible within the polymersome lumen (**Fig. 2d**). Further confirmation of effective encapsulation is shown in **Fig. S1**, where UV-Visible spectroscopy upon incubation of protein-loaded polymersomes with protease trypsin shows that the enzyme degrades non-encapsulated myoglobin within 4hrs (gradual disappearance of the Soret band, which

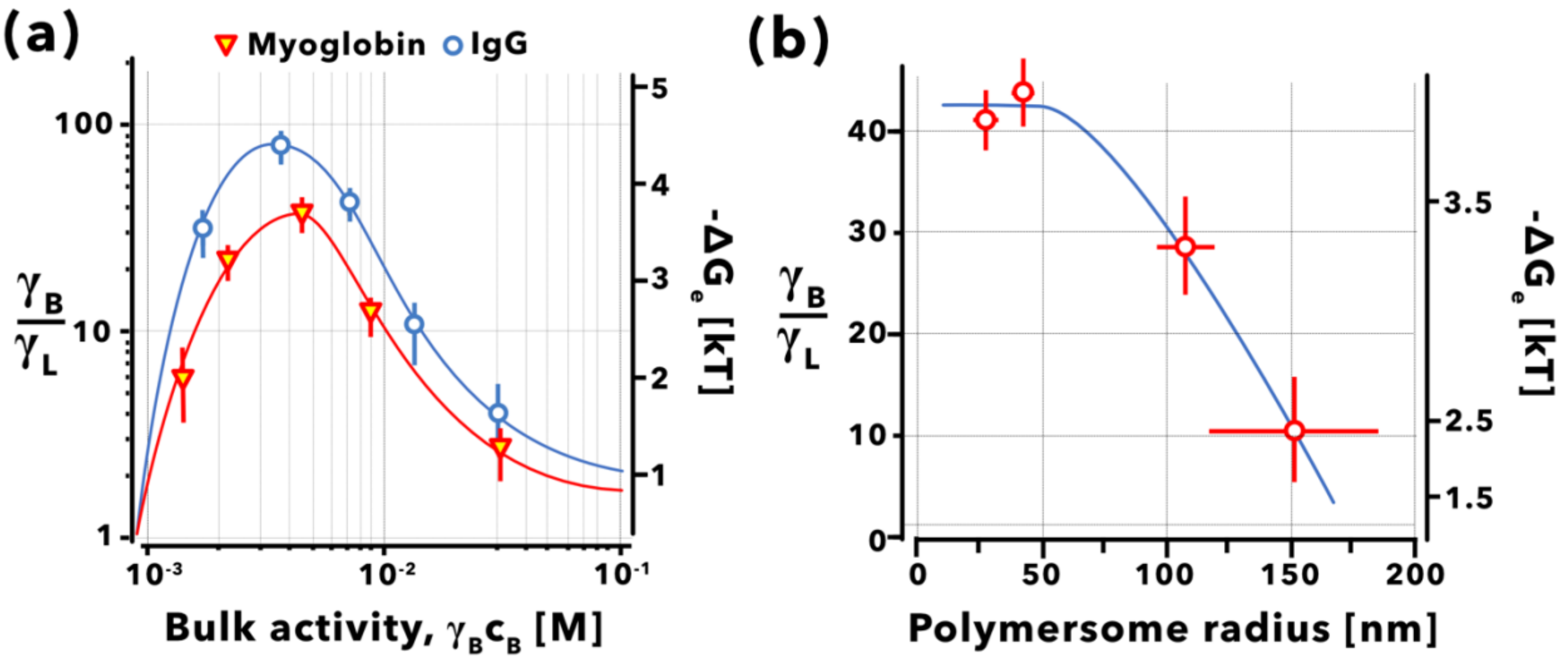

**Fig.3 (a)** Both myoglobin and IgG lumen and bulk myoglobin activity coefficient ratio and free energy of encapsulation as a function of their bulk activity (n = 5; error bars = ± SD). **(b)** Bulk and lumen myoglobin activity coefficient ratio and free energy of encapsulation as a function of the polymersome radius. (n = 5; error bars = ± SD).

is characteristic of non-degraded protein). In contrast, spectra recorded for myoglobin-loaded polymersomes remain identical to those obtained for the native protein, indicating that the trypsin cannot physically access the encapsulated myoglobin located within the interior of the polymersomes. Using the data in **Fig.2d** and the myoglobin protein molecular volume $v_p = 18.87$ nm$^3$ calculated using Chimera (PDB: 1WLA), we estimated the protein lumen volume fraction, $\phi_L$, as well as the corresponding Wigner–Seitz mean inter-particle distance calculated as:

$$< r_{ij} >= (\frac{6v_p}{\pi\phi_L})^{\frac{1}{3}} \qquad (2).$$

Both parameters, $\phi_L$, and $<r_{ij}>$ are plotted in **Fig.2g** as a function of the polymersome size showing that myoglobin proteins occupy an average 10% of the lumen volume in small polymersomes while the volume fraction decreases to 5-4% in lager polymersomes. In both cases, the mean inter-particle distance is quite short varying from 7 to 9 nm in the least dense configuration.

**The vesicle lumen as a different liquid-phase.** One constant and surprising result that emerges here, as well as in past reports (41-46), is that the concentration of the protein within the polymersome lumen, $c_L$ seem to be always higher than the protein bulk concentration $c_B$. However, such a counterintuitive result is the consequence of how we measure protein concentration. We use HPLC with the appropriate calibration to measure protein concentration, and to be more precise we measure the bulk protein activity $a_B = \gamma_B c_B$ where $\gamma_B$ is the coefficient of activity that indicates the deviation from ideality. The encapsulated protein is also measured in bulk as the polymersomes are broken down and the cargo is dispersed before the measurement. We thus define this to be the apparent lumen activity as a function of the lumen protein concentration, $a'_B = \gamma_B c_L$.

From a thermodynamic point of view, the process of encapsulation is driven by the difference between the protein chemical potential in bulk, $\mu_B = \mu_0 + kTln\gamma_B c_B$ and in the vesicle lumen $\mu_L = \mu_0 + kTln\gamma_L c_L$ with $\mu_0$ being the protein standard potential, $k$ the Boltzmann constant, $T$ the temperature and $\gamma_L$ the protein activity coefficient within the lumen. Assuming equilibrium conditions, $\gamma_B c_B = \gamma_L c_L$, we can then write

$$a'_B = \frac{\gamma_B}{\gamma_L} a_B \qquad (3).$$

The activity coefficients are a measure of the free energy of non-specific interaction between the proteins in water arising from self-excluded volume effects, electrostatic interactions and hydrophobic effect. The activity coefficient is thus related to the free energy change from ideal to a real solution and obeys the relation

$$kTln\gamma_i = G_i - G_i^{ideal} \quad (4),$$

where the $G_i$ is the protein free energy in the real solution and the $G_i^{ideal}$ is the protein free energy in an ideal solution. This latter is independent of whether the protein is placed within the bulk or the lumen. If we combine equation 3 and 4, we can thus calculate the free energy of encapsulation as

$$\Delta G_e = G_L - G_B = kTln\frac{\gamma_L}{\gamma_B} = kTln\frac{a_B}{a_B'} \quad (5).$$

Where $G_L$ and $G_B$ are the protein free energy in the lumens and bulk. In **Fig.3a**, we plot the ratio between the lumen and bulk activity coefficient for myoglobin and IgG measured loading the protein within polymersomes starting from different bulk activities. We also use equation 5 to calculate the corresponding free energy of encapsulation. The graph shows a bimodal trend with a peak at $\gamma_B/\gamma_L \approx 40$ and $\Delta G_e =$-3.71kT for myoglobin and $\gamma_B/\gamma_L \approx 80$ and $\Delta G_e =$-4.35kT for IgG and as the bulk activity increases $\gamma_B/\gamma_L \rightarrow 1$ while $\Delta G_e \rightarrow 0$. Albeit only a few kTs, the protein encapsulation within polymersomes is a favorable process indicating a stabilization of the mixture protein/water within the vesicle lumen. It is worth mentioning that the non-linear trend is expected and associated with the non-ideal nature of the protein-water solution. IgG, is greater in size than myoglobin and it does show a larger deviation from ideality. If we plot the $\gamma_B/\gamma_L$ ratio and $\Delta G_e$ for the myoglobin measured at the peak where $\gamma_B/\gamma_L \approx 40$ as a function of the polymersome radius, as shown in **Fig.3b**, we observed a decrease where for larger vesicles $\gamma_B/\gamma_L \approx 10$. Assuming a linear trend we can estimate that $\gamma_B/\gamma_L = 1$ corresponds to a polymersome radius of about 200nm. These results suggest that the equilibrium between the lumen and the bulk phases depends on the level of water confinement which seems to decrease with larger vesicles (see **Fig.2g**). This result indicates that as the vesicle gets smaller the ratio of interfacial/bulk water increases and consequently this affects the solubility of proteins which, as suggested by our data, increases leading to a more condensed phase.

**The vesicle encapsulation effect on protein folding-unfolding.** Myoglobin is a globular protein comprising 153 amino acids folding around a central HEME prosthetic group implicated in oxygen, NO, CO, and $H_2O_2$ storage. Myoglobin was the very first protein whose structure was resolved by Kendrew *et al.* (48), with several high-resolution crystal structures being reported (49). Several spectroscopic assays can assess the myoglobin secondary structure (50) making it one of the most studied globular protein. To study protein stability under vesicle confinement myoglobin-loaded polymersomes, as well as free myoglobin mixed with empty polymersomes, were heated from 30°C up to 95°C at pH 7.4 in a step-wise fashion. Samples were allowed to equilibrate for one hour at 5°C intervals to avoid hot spots and thermal gradients. Each sample was subsequently allowed to cool directly to 20°C, and once the sample reached final equilibrium temperature, the protein-loaded polymersomes were dissolved at pH 6 by addition of dilute HCl (0.01M) to release their protein payload. The resulting aqueous solutions were analyzed by both UV-Visible and fluorescence spectroscopy (50). The folded conformation of the native protein has a characteristic single Soret band at 410 nm. Upon denaturation, a second band due to protein unfolding appears at 390 nm (**Fig. 4a**) while it is clear that the encapsulated proteins show no apparent denaturation. Such an effect was observed in all the different polymersome sizes and all conditions of encapsulation as shown in the graph in **Fig. 4b** where the ratio between the absorbance at 410 and 390 nm are plotted. Similar results were observed using fluorescence spectroscopy, which monitors the emission due to the α-helix tryptophan (TRP7 and TRP14). After excitation at 295 nm, these aromatic groups exhibit a typical emission peak at 310 nm. When the protein unfolds, and its secondary structure is lost, the tryptophan moieties are exposed to the surrounding water. This degradation is associated with a second, more intense peak at 340 nm (see **Fig. 4c**). Tryptophan emission was not detected for any of the encapsulated protein samples independently of the polymersome size, further suggesting that polymersomes confer excellent protection against thermally induced denaturation (**Fig.4c-d**). Both UV-Visible and fluorescence spectroscopy were performed after the thermal denaturation cycle was applied to the systems and the polymersomes were dissolved at mild acidic pH. This means that even though the encapsulated myoglobin resisted the thermal treatment, the data in **Fig.4** do not discard a possible reversible unfolding favored by the confinement. In order to study the protein thermal denaturation during the thermal treatment, we performed in line circular dichroism (CD) studies on both free and encapsulated protein. As shown in **Fig.5a**, structural degradation of the native protein occurs between 70°C and 80°C and is complete at 95°C. When myoglobin is confined within polymersomes, the resulting CD spectra exhibit minimal temperature-dependent shift, and the ellipticity remains low even at 95°C (see **Fig.5b**). It is important to note that CD is rather insensitive to the scattering properties associated with polymersomes and CD spectra of empty polymersomes have zero ellipticity across all wavelengths (data not reported). This effect is not surprising as CD measures the differential absorption of left and right circularly polarized components of plane-polarized radiation and this is present only when the chromophore is either chiral or conjugated to a chiral center (51). Such insensitivity was also reported for both large and small lipid vesicles (52).

In **Fig.5c** we report the fractions of the folded protein, $f_{220}$ and $f_{208}$, measured at 220 and 208 nm respectively as a function of the temperature. These wavelengths are those associated with the characteristic peaks for α-helix (see methods for calculation). The data shows that the free protein has $f_{220}$ = 55% and $f_{208}$ = 48% of its structure folded and both decrease to 20% and 10% respectively at about 75°C in line with previous reports (51). The polymersome encapsulated protein folded fractions are always higher than the free protein with $f_{208}$ = 95% independently of the temperature and the $f_{220}$ dropping from 64% to 48%. Such a lack of unfolding can be explained by the high confinement effect that myoglobin experiences within the vesicles with a mean inter-particle distance of less than 10nm **Fig.2g**. Structurally the myoglobin unfolding will result in a considerable change in the radius of gyration with the unfolded structure being about twice the folded one as schematized in **Fig.5d**, where the PDB deposited structure (1wla) is compared with an unfolded one where the same amino acid sequence is minimized in theta solvent condition (i.e. $R_G \propto N_{AA}^{\frac{1}{2}}$).

**The vesicle encapsulation effect on protein structure.** The absorption in chiral peptide bonds in the UV region results from the amides double bond intense $\pi \rightarrow \pi*$ transition at c.a.190 nm and weaker and broader peak from the nitrogen electron pair $n \rightarrow \pi*$ transition at c.a. 220 nm. Depending on the peptide secondary structure and hence what angles the peptide bonds forms, the two transitions occur at different energies. For example, the CD spectrum for α-helix comes with two negative bands at 222 nm and 208 nm, β-sheet have a weak negative band at ~218 nm, and finally, a random coil shows a negative band at around 195 nm. Myoglobin structure (**Fig.5d**) comprises eight helices and, as shown in both **Fig.5a** and **Fig.6a**, its CD spectrum shows the α-helix distinctive bands at 208 and 222 nm respectively. However, for the encapsulated myoglobin at 20°C and 95°C the far UV CD spectra (**Fig.6a**) show a more negative ellipticity with a blue shift compared to the free protein. The sample incubated at 20°C shows that the two α-helix peaks shift from 222 to 218nm and from 208 to 205nm respectively with the latter increasing in intensity. The peak at 208 nm in the far UV spectra of the encapsulated protein at 20°C is consistent with the emergence of the $3_{10}$ helix arrangement (53), the third most common structural element observed in globular proteins. The $3_{10}$ helix has a different hydrogen-bonding pattern compared to the α-helix, with the carbonyl amide hydrogen bonds linking amino acids every three units rather than four (54).

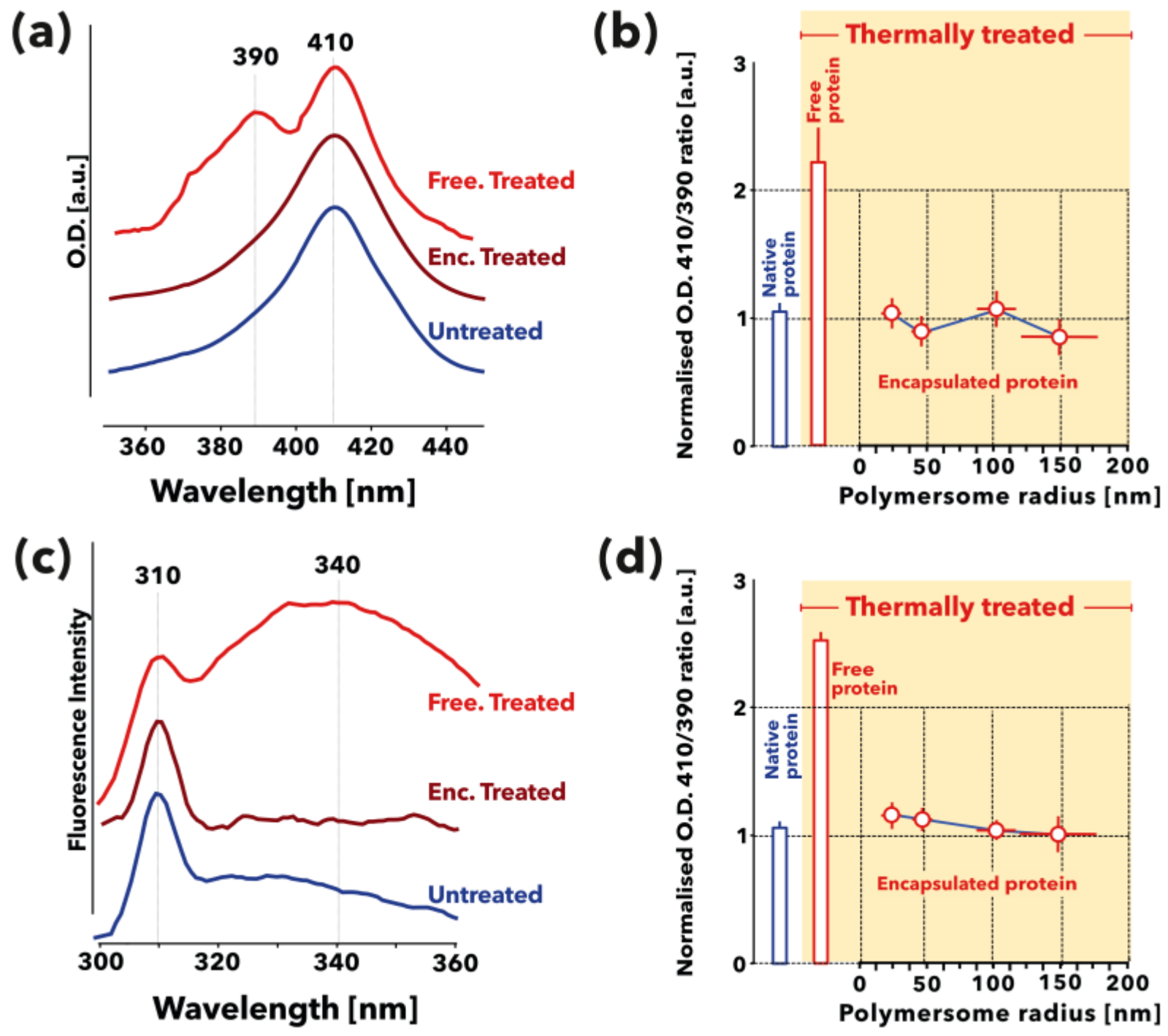

**Fig.4** **(a)** UV-visible spectra recorded for native untreated myoglobin, thermally-treated free and encapsulated myoglobin within 200 nm extruded polymersomes. **(b)** Normalised ratio between the absorbance intensity at 390 nm and 410 nm for the various encapsulation conditions (n = 3 error bars = ± SD). **(c)** Fluorescence spectra recorded for native myoglobin, thermally-treated free and encapsulated myoglobin within 200 nm extruded polymersomes. **(d)** Normalised ratio between the fluorescence intensity at 340 nm and 310 nm for the various encapsulation conditions (n = 3; error bars = ± SD).

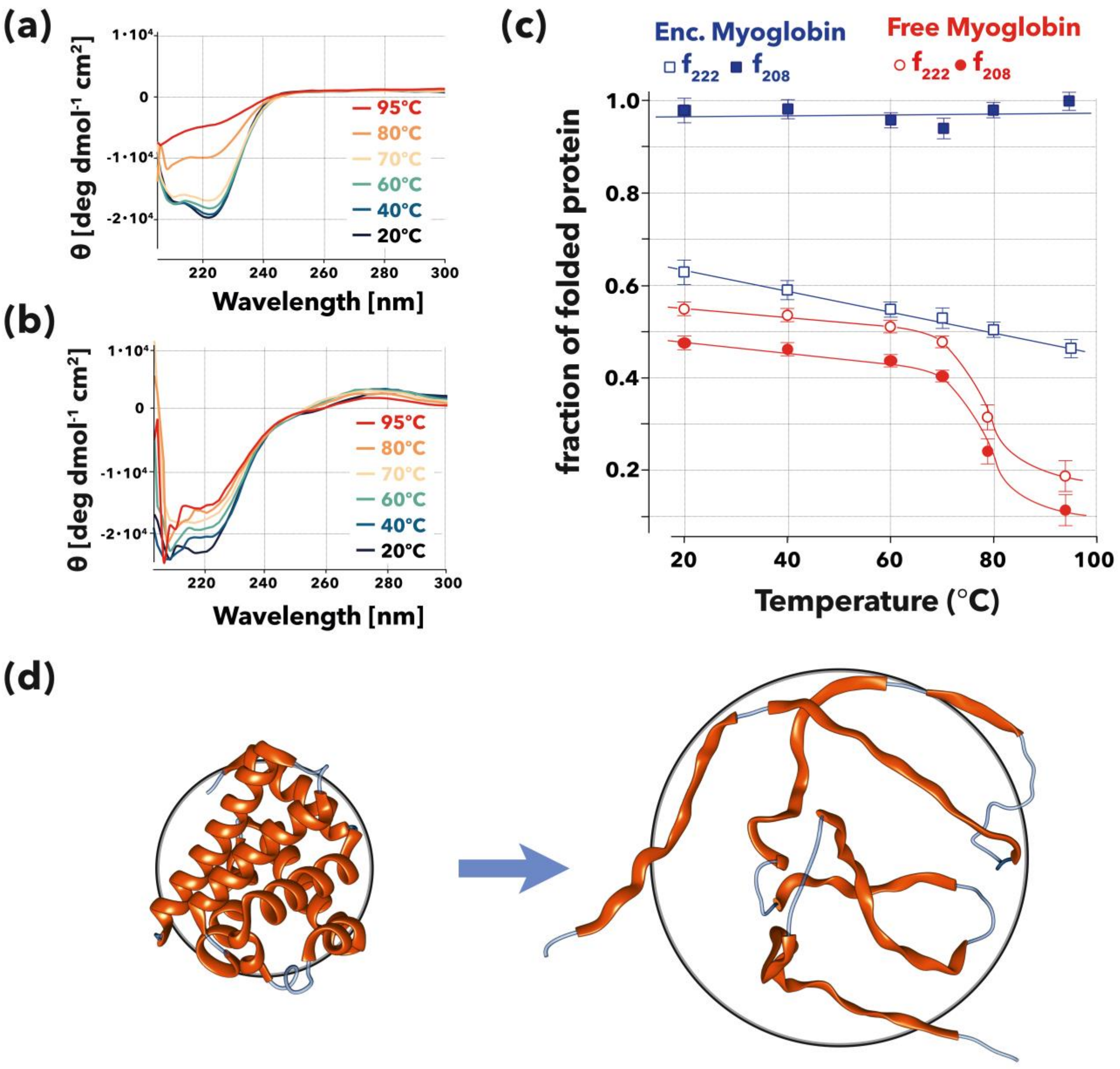

**Fig.5** Circular dichroism spectra recorded for free **(a)** and encapsulated **(b)** myoglobin as a function of the solution temperature. Note the polymersomes used have a 50nm radius. Fraction of folded protein **(c)** calculated at 208 and 222nm for both free and encapsulated myoglobin as a function of the solution temperature (n = 3; error bars = ± SD). Structure (PDB:1wla) of myoglobin undergoing unfolding into a random coil **(d)**. Note this latter was minimised in theta solvent condition (i.e. $R_G \sim N_{AA}^{0.5}$).

Furthermore, the two helices differ from one another by the dihedral angle that two consecutive residues make, with α having a 100° while in the $3_{10}$ forming a 120° angle around the helical axis. A considerable change of the myoglobin structure is also observed for the samples at 95°C, with the free protein sample showing the typical spectrum of random coils, while the encapsulated protein spectrum displays a negative

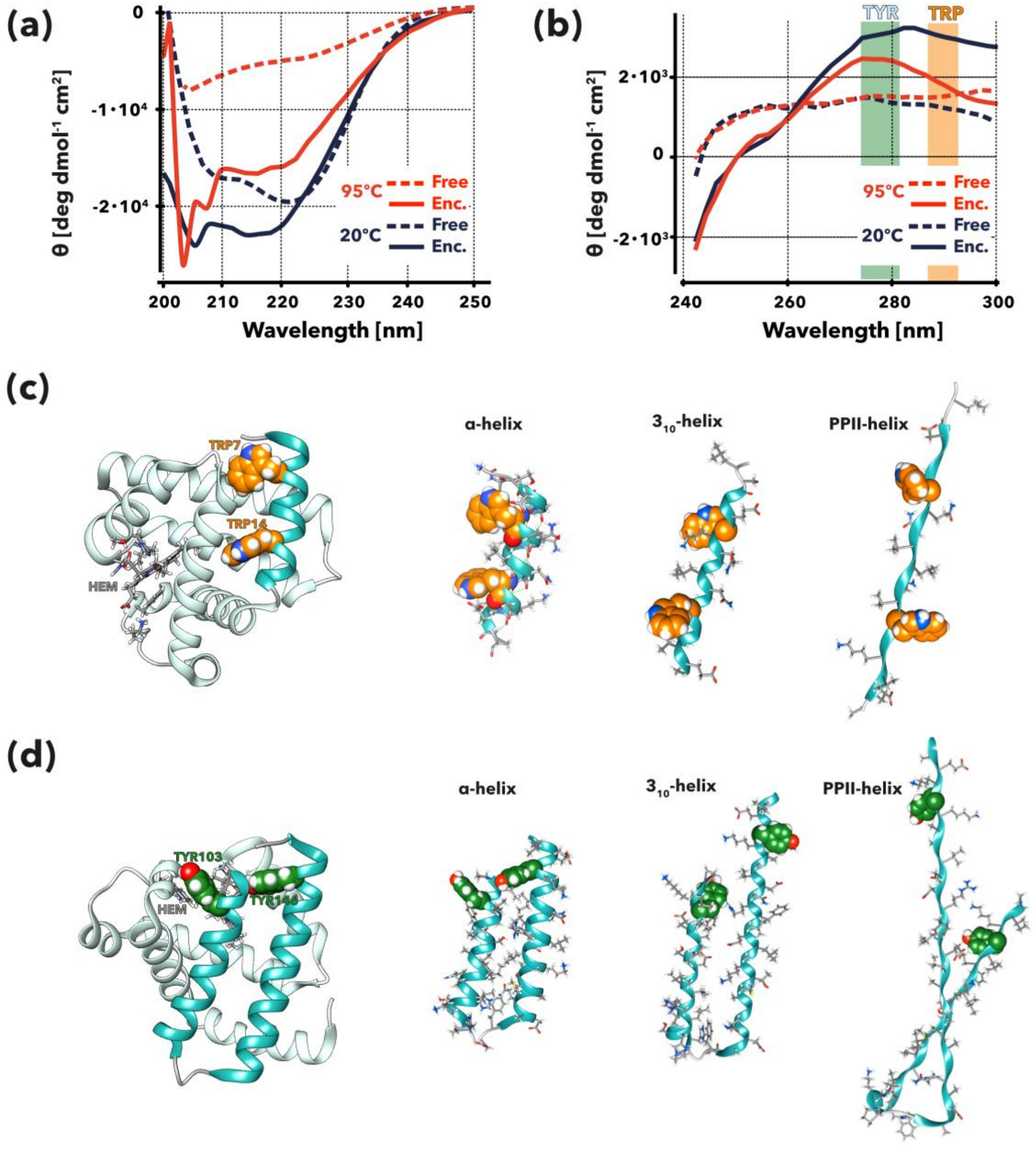

**Fig.6** Far **(a)** and near **(b)** UV Circular dichroism spectra recorded for free and encapsulated myoglobin at 20ºC and 95ºC. Note the polymersomes used have a 50nm radius. Myoglobin structure (PDB:1wla) with helix 1-20 residue carrying the two tryptophans in orange **(c)** and the two helices across 100-151 residues that bears the two tyrosines in green **(d)**. The two domains have been rearranged with three different helical arrangements, α, $3_{10}$ and PPII helix.

peak at 203 nm and a less negative peak at 208 nm with a further broader peak at 222 nm. At 95°C temperature is unlikely that the protein structure is controlled by hydrogen bonding, yet the negative peak and the percentage of folded structure suggest the presence of a secondary structure rather than a coiled unfolded configuration. Both aspects lead us to conclude that parts of the myoglobin sequence are folded into polyproline II (PPII) helices. Such a secondary structure is the dominant conformation in collagen and other fibrillar proteins (55). A characteristic feature of PPII is its repetitive torsional angles which form without any hydrogen bonds but only via interaction with water (55). Differences between the free and encapsulated proteins are also visible in the near UV CD spectra as in **Fig.6b** where encapsulated myoglobin at 20°C and 95°C spectra show an increase in ellipticity with a broad peak appearing between 270 and 300 nm. This region is associated with aromatic amino acids, with tyrosine (green band) peaking between 275 and 282nm and tryptophan (orange band) peaking at 290nm (51). Tyrosine and tryptophan are highlighted in green and orange respectively in the myoglobin structures shown in **Figs.6c-d** where we assessed the possibility of helical arrangement changes. We thus isolated the helix from 1 to 20 residue carrying the two tryptophans (**Fig.6c**) and the two helices spanning from 100 to 151 residues that bears the two tyrosines (**Fig.6d**). In both cases, we rearrange the helical arrangement imposing both $3_{10}$ and PPII dihedral angles, and the respective structures are shown in **Figs.6c-d**. A potential conformational change from α to $3_{10}$ and to PPII forces the aromatic residues farther apart which otherwise would be within the 1nm distance where the aromatic group absorbance will interfere with each other (51). The interplay between the different helical arrangements was also postulated during folding-unfolding transitions (56), and indeed the different structures seem to co-exist at the same time in protein structures (54) (57). We thus propose that the different thermodynamic state encountered within the lumen favors the formation for the $3_{10}$ helix arrangements and hence it is responsible for the myoglobin to present a more condensed arrangement. At higher temperatures, the loss of hydrogen bonding drives the full unfolding, and completely loss of any secondary structure on the myoglobin when it is free, but when it is encapsulated this seems to evolve differently and a considerable portion of the amino acids maintain a helical arrangement albeit being more consistent with the PPII conformation

**Enzyme activity within vesicles.** The structural stability of the studied polymersome-confined protein is further confirmed by monitoring its enzymatic activity after applying the thermal cycle. The enzymatic activity can be assessed by measuring the rate of oxidation of guaiacol into its tetramer using UV-Visible spectroscopy at pH 7.4 and 25°C (all measurements were normalized so as to have equal protein substrates concentrations). The HEME group that is responsible for myoglobin's enzymatic activity is susceptible to denaturation, and expectedly the free protein is no longer enzymatically active after the thermal treatment as shown in **Fig. 7a**. Oppositely, the protein is still active in all vesicles confirming that the activity of the encapsulated myoglobin HEME group is still held in a reactive configuration even after thermal treatment. Interestingly, the activity of protein encapsulated within the polymersomes increases concerning the native free protein **Fig. 7b**. Although within the experimental error, the augmentation is higher within the smaller vesicles where we measured almost twice the activity compared to the free protein (considered 1 here) and only 1.5 for larger vesicles. Enhanced enzymatic activity as a function of macromolecular crowding has already been reported for DNA polymerase (58), multi-copper oxidase (59), and ribozyme (60). However, most of these studies utilized inert crowding agents, whereas, in the present work, myoglobin activity appears to exhibit an auto-catalytic effect once encapsulated within polymersomes under relatively crowded conditions. A similar phenomenon was observed by de Souza et al., who reported that encapsulating the entire ribosomal machinery inside 100 nm lipid vesicles produced an average yield of fluorescent protein more than six times higher than that found in bulk water (61).

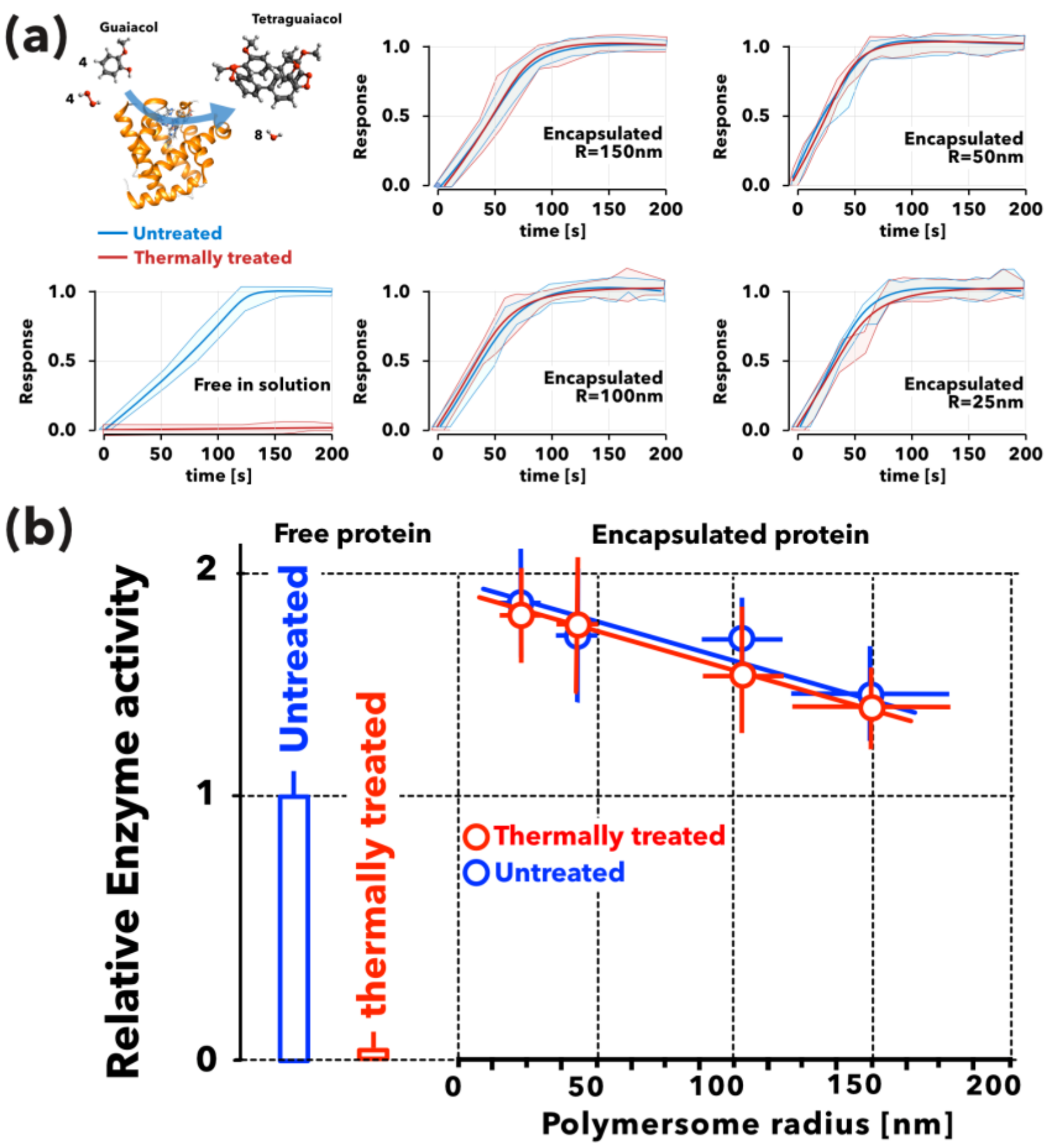

**Fig.7** Schematic of the oxidation of guaiacol into its tetramer catalysed by the myoglobin **(a)**. The reaction can be monitored using UV-Visible spectroscopy and the free and encapsulated protein normalised responses are plotted for the native and thermally treated samples. (Note the reactions were performed at pH 7.4 and 25°C and all measurements were normalised so as to have equal protein and substrates concentration. The corresponding enzyme activity **(b)** was normalised to the free sample (1) and plotted as a function of the polymersome size for both treated and untreated samples (n = 3; error bars = ± SD).

## Conclusions

The data shown herein demonstrate a valid platform for studying protein dynamics under conditions that closely resemble those found in vivo. We show that vesicle entrapment leads to a considerable change on the protein/water interaction creating a more condensed phase with a consequent higher density of protein per volume compared to bulk conditions. In many ways, this suggests that the protein/water mixture within the lumen is in a different liquid-state than the bulk mixture. In this fashion, the protein encapsulation process corresponds with a liquid-liquid phase transition wherein the lumen phase is more condensed than in the bulk phase. The marked enhancement in thermal stability, as well as the enhanced enzymatic activity for the encapsulated myoglobin, corroborate the different energetic scenario furthermore. The stabilization effect observed across a range of polymersome diameters suggests that confinement and crowding are intimately connected. Both phenomena lead to the formation of a highly confined water network between the encapsulated proteins and the polymersome inner leaflet, with the water molecules forced to occupy nanoscopic volumes ranging in size from a few nm to tens of nm. Several studies on water confined between two hydrophilic substrates reported a more glass-like structure with disrupted hydrogen bonds, exhibiting somewhat longer lifetimes than the typical picoseconds, together with a reduction in tetrahedral bonding arrangements (62, 63). Other studies have shown that this interfacial effect can extend well beyond the electrical double layer thickness of a few nm and reach the µm range with the concomitant formation of a less flexible, more organized phase (64-66). More relevantly to the present case, Bhattacharyya et al. measured water solvation dynamics using fluorescent probes within vesicles and observed two regimes: one very slow (> 1 ns) and one faster ( ~600ps). Both relaxation times are considerably slower than bulk water ( >1ps) and attributed to the interface-bound water and the water within the vesicle lumens (67-69).

In conclusion, we provide strong evidence that protein thermodynamics within a nanoscopic aqueous environment is strongly affected by both protein concentration and spatial confinement. More importantly, we demonstrate that proteins encapsulated within polymersomes can withstand large temperature gradients without compromising their structure (and hence their biochemical activity). We believe that such a finding is significant in the context of polymersome-mediated delivery of proteins and the development of nano-reactors. Finally, our findings suggest a new perspective for the 'origin of life' research, as we propose a new paradigm for compartmentalization. We demonstrate that compartmentalization is not just critical for the spatial separation of aqueous volumes, but it also offers a potentially important stabilization mechanism for proteins, which are one of life's essential building blocks.

## Acknowledgements

We would like to thank the Human Frontiers Science Program (RGY0064/2010 ) for funding DC salary, the ERC (ERC-MEViC-278793, ERC-CheSSTaG 769798), for funding part of GB and LRP salaries, BTG ltd for funding LC stipend, the EPSRC grant EP/E03103X/1 for funding CLP salary, grant EP/G062137/1 for funding XT salary, grant EP/N026322/1 for part of GB and AP salary. All of them for funding the research and access to facilities associated with this project.

## Supporting information

**Materials.** Myoglobin from equine skeletal muscle lyophilized powder, (Sigma), 2-methoxyphenol (Guaiacol), (Sigma), hydrogen peroxide 30%, (Sigma), Trypsin, (Sigma), Trypsin inhibitor: Complete protease inhibitor cocktail tablets, (Roche) and Immunoglobulin G (IgG) (I5006; Sigma-Aldrich, UK) and 5 nm AuNP-IgG (gold labelled immunoglobulin G) (Sigma-Aldrich, UK), Cy5 (Cyanine5) NHS ester dye (Lumiprobe) dimethyl sulfoxide (DMSO), methanol and chlorform (Sigma-Aldrich, UK), Sephadex G-25 (Sigma-Aldrich, UK).

**IgG dye labelling.** In the first step the proteins were solubilised in pH ~8.3 sodium bicarbonate buffered solution at a final concentration of 5 mg/ml. Subsequently, 0.25 mg of (λex= 649 nm, and λem= 669 nm) previously dissolved in 20 µl DMSO were added to the protein solutions and incubated at room temperature for 2h in stirring conditions. This was necessary to allow the protein-dye conjugations between the succinimidyl ester group present on the dye chemical structure and primary amines presents on the proteins. The labelled proteins were thus purified from the unbounded dye via SEC using a Sephadex G-25. Finally, the absorbance at 280 nm of the purified product was measured. The absorbance of free dye was subtracted from the total absorbance and protein concentration was determined against a standard curve.

**$PMPC_{25}$-$PDPA_{70}$ polymersomes preparation and myoglobin encapsulation.** $PMPC_{25}$–$PDPA_{70}$ copolymers were synthesized by atom transfer radical polymerization (ATRP), as reported elsewhere[1]. In a typical experiment, $PMPC_{25}$–$PDPA_{70}$ powder was dissolved using Methanol/Chlorform 1:2 and a thin film was casted on a glass vial. rehydrated by adding 2 ml of 1X phosphate buffered saline (PBS 0.1 M) at pH 7.4. The buffered solution was stirred (magnetic stirring at 200 rpm) for 8 weeks to allow polymersomes formation. Both IgG and Myoglobin encapsulation by electroporation was tested using different conditions such as different initial protein concentrations with 5 pulses, 2500 V each pulse. After mixing proteins and polymersomes at the desired conditions depending on the experiment, 400 µl of the mixture was loaded into a 2 mm width gap electroporation cuvette (Eppendorf, UK) and electroporated using an Electroporator 2510 (Eppendorf, UK) instrument, applying a voltage of 2500 V each pulse.

**Polymersomes extrusion and encapsulation efficiency.** Polymeric vesicles were extruded using a Liposofast extruder. 50, 100, 200, 400 nm pore sized polycarbonate membranes were used to obtain polymersomes of the correct size, loaded with myoglobin. Polymersomes were then purified via gel permeation chromatography (GPC), using Sepharose 4B as stationary phase and PBS at pH 7.4 was used to elute the polymersomes. Proteins per vesicles were determined by Reversed-phase high pressure liquid chromatography (RP-HPLC).
An anti-tubulin IgG conjugated with AlexaFluor® 647 dye (λex= 650 nm, and λem= 669 nm) (ab6161; Abcam®, UK) was used as a model. Myglobin was used pristine and measured by UV/Vis absorbance at 408nm. The calibration curves were obtained using an RP-HPLC (Dionex, Ultimate 3000) with a C18 analytic column (Phenomenex® Jupiter C18, 300A, 150 x 4.60 mm, 5 micron) and using a constant flow ratio of 1 ml/min. The eluents used were milliQ H2O added of 0.05 % V/V trifluoroacetic acid (TFA) (eluent A) and CH3OH with 0.05 % TFA (eluent B) mixed according a gradient starting at 20% increasing at 40% after 6mins elution time, 45% after 16mins elution time 50% after 20mins, 70% after 23mins to peak at 100% after 24mins. After 27mins of elution time the gradient was decreased to 20% and maintain at this value for the rest of the analysis.

**Dynamic Light Scattering and transmission electron microscopy**. Dynamic light scattering measurements were performed using the Malvern Zetasizer Nano set at 20 0C. Samples were diluted to 0.15 mg/ml in 1X PBS pH 7.4. 800 µl of diluted sample was then placed into a polystyrene cuvette (Malvern, DTS0012) and analyzed. TEM analyses were performed using a FEI Tecnica Spirit microscope with maximal working voltage of 120 KV and equipped with Gatan1K MS600CW CCD camera. The copper grids used for the sample's analysis were initially coated with a carbon layer (thickness: ~20 nm) using a carbon coater. Subsequently, the prepared grids were submerged in the polymersomes and afterwards stained using a phosphotungstic acid (PTA) solution (0.75% w/w) as described in previous works (Pearson et al., 2013; Wang et al., 2012). The PTA staining was herein applied since it enables the detections of the ester bonds presents in the PMPC- PDPA molecular structure.

**Trypsin stability.** Polymersomes loaded with myoglobin were incubated at 37°C in 0.1 M PBS, pH 7.4 at 10 µg/ml of protein together with trypsin. The trypsin/myoglobin molar ratio was 1:2. In the same conditions empty polymersomes were used as control. UV-Vis spectra of the encapsulated myoglobin were recorded immediately after the addition of trypsin ($t_0$), after 4h and 24 hours of incubation. In order to remove the scattering of the polymersomes the same approach described before for evaluating the EE was followed, i.e. solubilization of polymersomes. In this case, though, protease inhibitor was added at a 5%(v/v) (concentration) before solubilizing the polymersomes to prevent trypsin degradation of the released myoglobin that would alter the results. Measurements were performed in triplicate. The UV-Vis spectra were recorded in the 800-200nm measurement range to follow the changes in the peak at 408nm.

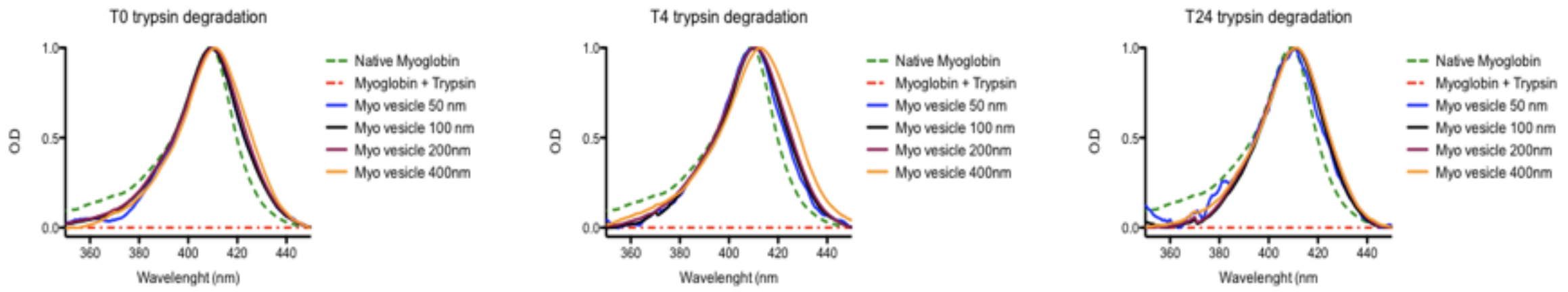

**Figure S1** Trypsin degradation effect on polymersomes at different sizes and free myoglobin compared with native myoglobin (green). The UV-Vis spectra were measured in the range 800-200nm.

**Thermal stability.** Myoglobin polymersomes with different diameters in 0.1 M PBS, pH 7.4, were incubated into a thermostated Peltier chamber where the temperature was gradually raised from 30 to 95°C with an interval step of 5°C and a concentration of 10 µg/ml of protein. Each temperature was maintained for 60 minutes. The polymersomes preparations were then solubilized at pH 6 and the UV-Vis spectra 800-200 nm were recorded. Empty polymersomes where also used as control.

**Myoglobin secondary structure analysis.** Circular dichroism (CD) measurements of the polymersomes at 10 µg/ml of protein, were recorded at different temperatures 20, 40, 60, 70, 80, 95°C after placing the samples in a thermostated chamber where the temperature was increased with ramps of 10°C. Each temperature was maintained for 60 minutes. The changes in the far-UV CD spectra and determination of ellipticity at 222 nm were analyzed.

**Myoglobin tertiary structure analysis.** The tryptophan fluorescence spectra 300nm to 370nm of the myoglobin polymersomes at different diameters in 0.1 M PBS, pH 7.4, were recorded at a concentration of 10 µg/ml of myoglobin. The excitation was set at 295nm. The increase of fluorescence correlated with protein denaturation was measured before and after temperature denaturation. Results were normalised as ratio of 310/340 nm pick intensity.

**Myoglobin polymersomes bioactivity after thermal denaturation.** Bioactivity was measured for all polymersome samples after the thermal denaturation ramp from 30 to 70°C by using a thermostated chamber. 2-methoxyphenol (Guaicol) and $H_2O_2$ were added at concentration of 3µM and 0.4mM respectively. Activities were measured as the increase in the formation at 470 nm of tetraguaiacol with the time. Values are showed as normalized increase of absorbance and the activity is plotted normalized to the free enzyme as unit.

**Molecular graphics and analysis.** We use the protein database (PDB) 1IGY for IgG and 1wla for myoglobin. All analyses performed with UCSF Chimera, developed by the Resource for Biocomputing, Visualization, and Informatics at the University of California, San Francisco, with support from NIH P41-GM103311.